\documentstyle[twoside,fleqn,npb,axodraw]{article}
\newcommand{\AmS}{{\protect\the\textfont2
  A\kern-.1667em\lower.5ex\hbox{M}\kern-.125emS}}

\newskip\humongous \humongous=0pt plus 1000pt minus 1000pt

\newif\ifdtup

\newcounter{eqnumber}[section]

\def\ksl{\not{\hbox{\kern-2.3pt $k$}}}

\def\e{\epsilon}
\def\musq{\mu^2}

\def\eps{\epsilon}

\def\Ord{{\cal O}}
\def\cm{{\cal M}}

\def\Nf{{N_{\! f}}}

\def\MSbar{\overline{\rm MS}}

\def\li#1{{\mathop{\rm Li}\nolimits}_#1}
\def\Li{\mathop{\rm Li}\nolimits}

\def\Sublead{{\rm \scriptscriptstyle SL}}
\def\Lead{{\rm \scriptscriptstyle L}}
\def\alphas{\alpha_s}

\def\eqn#1{eq.~(\ref{#1})}

\def\spa#1.#2{\left\langle#1\,#2\right\rangle}
\def\spb#1.#2{\left[#1\,#2\right]}
\def\lor#1.#2{\left(#1\,#2\right)}

\title{$\null$\\
 \vskip - 1.6 cm  
{\small DESY 02-202 \hfill UCLA/02/TEP/35 \hfill SLAC--PUB--9578 } \\
 $\null$ \\
Two-loop corrections to $ gg \rightarrow \gamma \gamma $}

\author{Z. Bern,\address{Department of Physics and Astronomy, UCLA,
                        Los Angeles, CA 90095-1547, USA}%
   \thanks{Research supported by the US Department of Energy under grant
          DE-FG03-91ER40662.} 
       A. De Freitas,\address{Deutsches Elektronen Synchrotron, DESY,
                               D-15738 Zeuthen, Germany}%
        \thanks{Presented at RADCOR 2002/Loops and Legs in Quantum Field Theory
                (September 2002, Kloster Banz, Germany).} 
        L. Dixon\address{Stanford Linear Accelerator Center,
                          Stanford University, Stanford, CA 94309, USA}%
      \thanks{Research supported by the US Department of Energy 
             under contract DE-AC03-76SF00515.}}

\begin{document}

\begin{abstract}
An overview of the calculation of the two-loop helicity
amplitudes for scattering of two gluons into two photons is presented.
These matrix elements enter into the recent improved calculation
of the QCD background to Higgs boson decay into a pair of photons, which is
the preferred search mode at the LHC for the case of a light Higgs boson.
\end{abstract}

\maketitle




Recent years have seen an enormous improvement in our ability to
calculate two-loop amplitudes with more than a single kinematic
invariant.  The initial calculations of this type were for four-point
scattering in maximally supersymmetric theories~\cite{BRY} and special
helicity configurations in QCD~\cite{AllPlusTwo}.  In this talk we
will focus on the two-loop amplitudes for gluon fusion into two
photons~\cite{gggamgam}.  This calculation is among the more general
two-loop
processes~\cite{BhabhaTwoLoop,GOTY2to2,GOTYgggg,lbyl,gggg,SUSYglover,%
SUSYpaper,EEto3jets} that have become doable, thanks to the
development of rather general two-loop integration
techniques~\cite{Integrals}.  Further summaries of these 
developments may be found in a number of talks at this
conference~\cite{OtherTalks}.

Gluon fusion into two photons is phenomenologically interesting
because the preferred mode for discovering a light Higgs boson ($M_H <
140$ GeV) at the LHC is through its decay into two photons.  There are
large backgrounds to this decay coming from radiation from either
partons or hadrons~\cite{TwoPhotonBkgd1,PiBkgd}.  Although the gluon
fusion contributions to the background are formally of higher order in
the perturbative expansion they are greatly enhanced by the large
gluon distribution at small $x$ at the LHC.

The amplitudes described here were calculated in the helicity
formalism, using the 't Hooft-Veltman dimensional regularization
scheme \cite{HV}.  Helicity methods have a long history, having been used
rather productively at tree and one-loop levels (see {\it e.g.}
refs.~\cite{SpinorHelicity,MPReview}).  At two loops the first of the
$2 \rightarrow 2$ amplitudes were also evaluated using helicity
states~\cite{BRY,AllPlusTwo}.  In the $gg \to \gamma \gamma$
process under consideration here, the tree amplitudes vanish and
the one-loop amplitudes give the leading order contributions.  Thus
the  next-to-leading order contributions to gluon fusion require
an interference of two-loop amplitudes with one-loop
amplitudes.  Instead of evaluating this interference directly, we again 
make use of helicity.

To generate the loop momentum integrals, we did not use Feynman
diagrams, but instead used a unitarity-based
technique~\cite{CutBased,BernMorgan}.  This technique exploits a
duality between loop and phase space integrals and has already been
employed in a number of two-loop calculations
\cite{BRY,AllPlusTwo,SUSYpaper}.  This duality was also used very
recently to calculate the exact NNLO contribution to the total cross
section for Higgs boson production at the LHC \cite{anamel}.

In performing the loop momentum integrals some minor extensions of the
general reduction algorithms developed for four-point massless
integrals~\cite{Integrals} were needed.  Any four-dimensional
polarization vector can be expanded in terms of the three independent
momenta in the problem plus a dual vector.  This means that dot
products of polarization vectors with loop momenta can be re-expressed
in terms of dot products of loop momenta with external momenta and also
the dual vector.  Integrals where no dual vector appears can be
reduced using the previously constructed reduction algorithms.  This
leaves integrals containing dual vectors dotted into loop momenta to
be evaluated; as discussed in ref.~\cite{gggg} such integrals
tend to be simpler to evaluate. An alternative method for dealing with
helicity at two loops was recently given in ref.~\cite{SUSYglover}.

We take the quarks in the loops to be massless since the mass of the
Higgs boson, and therefore the energy scale of the experiments for
which this calculation is relevant, is well above the mass of all
quarks other than the top.  Moreover, the top quark can be ignored
since the aforementioned energy scale is well below the $2m_t=350$ GeV
threshold.

The renormalized  $gg \to \gamma \gamma$ amplitude may be expanded as
\begin{eqnarray}
\cm_{gg \to \gamma \gamma} \hskip -.2 cm 
& =  & \hskip -.2 cm 4\pi\alpha \, 
\Biggl[
 { \alphas(\mu^2) \over 2\pi } 
\cm_{gg \to \gamma \gamma}^{(1)} \nonumber \\
&& \null 
+ \biggl( { \alphas(\mu^2) \over 2\pi } \biggr)^2
 \cm_{gg \to \gamma \gamma}^{(2)}
+ \cdots
\Biggr] \,, \nonumber
\end{eqnarray}
where $ \cm_{gg \to \gamma \gamma}^{(L)}$ is the $L$th
loop contribution, $\alpha_s(\mu^2)$ is the $\MSbar$ running QCD coupling
and $\alpha$ is the QED fine structure constant.

We use Catani's `Magic Formula'~\cite{Catani} for two-loop infrared
divergences to organize our results.  This formula was obtained for
general QCD amplitudes, but with minor modifications it also
applies to mixed QED and QCD amplitudes. In cases where
the tree-level amplitude vanishes (as happens for $gg \rightarrow
\gamma \gamma$), Catani's formula reduces to,
\begin{equation}
{\cal M}_{ {gg \to \gamma\gamma}}^{(2)} 
       = I^{(1)}_{gg \to \gamma\gamma}
        (\e) {\cal M}_{{gg \to \gamma\gamma}}^{(1)}
 + {\cal M}_{ {gg \to \gamma\gamma}}^{(2)\rm fin} \,,
\label{FiniteInfinite} 
\end{equation}
where $I^{(1)}(\e)$ contains the infrared singularities, and ${\cal
M}^{(2)\rm fin}$ is a finite remainder.  In our case the color factors
can only be proportional to $\delta^{a_1 a_2}$, so $I^{(1)}(\e)$ is
relatively simple:
\begin{eqnarray}
&& \hskip -7 mm 
I^{(1)}_{gg \to \gamma\gamma} (\e,\mu^2;\{p\}) =
\nonumber \\
&& \hskip -5 mm
-  N {e^{-\e \psi(1)} \over \Gamma(1-\e)} 
\biggl[ {1 \over \e^2}   
+  \biggl( {11\over 6} 
              - {1\over 3} {N_f\over N} \biggr) {1 \over \e}  \biggr]
\biggl({\mu^2\over -s} \biggr)^{\e}  \,, \nonumber
\end{eqnarray}
where $N=3$ in QCD and $N_{\! f}$ is the number of light flavors.  In
this formula the infrared divergences are encoded as poles in the
dimensional regularization parameter $\e = (4-D)/2$.

The one-loop amplitudes were given in ref.~\cite{gggamgam} through
their relation to the one-loop four-gluon amplitudes. 
We can write
\[
{\cal M}^{(1)}_{gg \rightarrow \gamma \gamma} = 
2 \, \delta^{a_1a_2}\biggl( \sum_{i=1}^{N_f}Q_i^2 \biggr) M^{(1)} \,,
\nonumber
\]
where $Q_i$ is the electric charge of the $i$th quark and
 $M^{(1)}$ is equal to the sum of non-cyclic permutations of the fermionic
contributions to the four-gluon primitive amplitudes~\cite{BernMorgan}. 

It is convenient to extract overall spinor phases from each helicity
amplitude,
\[
M^{(1)}(1^{\lambda_1}, 2^{\lambda_2}, 3^{\lambda_3}, 4^{\lambda_4}) = 
S_{\lambda_1 \lambda_2 \lambda_3 \lambda_4} 
M^{(1)}_{\lambda_1 \lambda_2 \lambda_3 \lambda_4} , 
\]
where 
\begin{eqnarray}
&& \hskip -7 mm
S_{++++} = i{\spb1.2 \spb3.4 \over \spa1.2 \spa3.4} \,, \hskip 3 mm 
S_{--++} = i{\spa1.2 \spb3.4 \over \spb1.2 \spa3.4} \,,
   \nonumber \\
&& \hskip -7 mm 
S_{-+++}  =  i{\spa1.2 \spa1.4 \spb2.4 \over \spa3.4 \spa2.3 \spa2.4} \,.
\nonumber
\end{eqnarray}

Through $\Ord(\eps^0)$ the amplitudes simplify greatly and the
$M_{\lambda_1\lambda_2\lambda_3\lambda_4}^{(1)}$ reduce to, 
\begin{eqnarray}
&& \hskip -7 mm
M_{++++}^{(1)} =
   1 + \Ord(\e)
\,, \nonumber \\
&& \hskip -7 mm
M_{-+++}^{(1)} = M_{+-++}^{(1)} = M_{++-+}^{(1)} = M_{+++-}^{(1)} \nonumber \\
&& \hskip -7 mm
\hphantom{M_{-+++}^{(1)}} =  1 + \Ord(\e), 
 \nonumber \\
&& \hskip -7 mm
M_{--++}^{(1)} =
- {1\over2} {t^2+u^2\over s^2}
                  \Bigl[ \ln^2\Bigl({t\over u}\Bigr) +\pi^2 \Bigr] \nonumber \\
&& \hskip -7 mm
\hphantom{M_{--++}^{(1)}}             
- {t-u\over s} \ln\Bigl({t\over u}\Bigr) - 1 + \Ord(\e)
\,,  \nonumber \\
&& \hskip -7 mm
M_{-+-+}^{(1)} =
 - {1\over2} {t^2+s^2 \over u^2}
                  \ln^2\Bigl(-{t\over s}\Bigr) \nonumber \\
&& \hskip -7 mm
\hphantom{M_{-+-+}^{(1)}}
    - i \pi \biggl[ {t^2+s^2\over u^2} \ln\Bigl(-{t\over s}\Bigr)
                      + {t-s\over u} \biggr]  \nonumber \\
&& \hskip -7 mm
\hphantom{M_{-+-+}^{(1)}}
 - {t-s\over u} \ln\Bigl(-{t\over s}\Bigr) - 1
 + \Ord(\e)
\,, \nonumber \\
&& \hskip -7 mm
M_{+--+}^{(1)}(s,t,u) =
M_{-+-+}^{(1)}(s,u,t)
\,.
\nonumber
\end{eqnarray}
where we are using an ``all-outgoing'' convention for the momentum
($p_i$) and helicity ($\lambda_i$) labeling.  The Mandelstam variables
are $s = (p_1+p_2)^2$, $t = (p_1+p_4)^2$, and $u = (p_1+p_3)^2 $.

We consider both QCD corrections with internal gluon lines and QED
corrections with internal photons. For the QCD corrections, the
dependence of the finite remainder in \eqn{FiniteInfinite} on quark
charges, $N$, $\Nf$ and the renormalization scale $\mu$, may be
extracted as,
\begin{eqnarray}
&& \hskip -7 mm
{\cal M}^{(2) \rm fin}_{gg \to \gamma\gamma}  = 
2 \, \delta^{a_1a_2} \, 
  \, \biggl( \sum_{j = 1}^\Nf Q_j^2 \biggr) \,
S_{\lambda_1 \lambda_2 \lambda_3 \lambda_4} \, \nonumber \\
&& 
\null \times \biggl[
   {11 N - 2 \Nf \over 6} \Bigl( \ln(\musq/s) + i \pi \Bigr) 
    M^{(1)}_{\lambda_1 \lambda_2 \lambda_3 \lambda_4}
\nonumber \\ 
&& 
 \null  + N F^\Lead_{\lambda_1 \lambda_2 \lambda_3 \lambda_4} 
  - {1\over N} \, 
      F^\Sublead_{\lambda_1 \lambda_2 \lambda_3 \lambda_4} \biggr] \,.
\label{FiniteRemainder}
\end{eqnarray}
The two-loop renormalized QED corrections are a little simpler, since in this
case the amplitudes are free of infrared divergences,
\begin{eqnarray}
&& \hskip -7 mm
{\cal M}^{(2){\rm QED}}_{gg \rightarrow \gamma \gamma} = 
4\, \delta^{a_1a_2} \biggl(\sum_{j=1}^{N_{\!f}}Q^4_j\biggr) 
 \nonumber \\
 && \hskip 12 mm
 \times \, 
S_{\lambda_1 \lambda_2 \lambda_3 \lambda_4}
F^\Sublead_{\lambda_1 \lambda_2 \lambda_3 \lambda_4} \,.
\label{QEDres}
\end{eqnarray}

We quote our results in the physical $s$-channel 
$(s > 0; \; t, \, u < 0)$.
In order to reduce the size of the expressions we define
\begin{eqnarray}
&& \hskip -7 mm
   x = {t\over s} , \quad 
   y = {u\over s} , \quad 
   X = \ln(-x) , \quad 
   Y = \ln(-y) , \nonumber \\
&& \hskip -7 mm
   {\tilde X} = X + i\pi , \quad  
   {\tilde Y} = Y + i\pi , \nonumber \\
&& \hskip -7 mm 
   \Xi = {\tilde X}^2 +\pi^2 , \quad
   \Upsilon = {\tilde Y}^2 +\pi^2 , \nonumber \\
&& \hskip -7 mm
    Z_{\pm} = X \pm Y , \quad
   {\tilde Z} = (X-Y)^2 + \pi^2 , \nonumber \\
&& \hskip -7 mm
   A^{\pm}_n = \Li_n(-x) \pm \zeta_n , \quad
   B = \li2(-x) - {\pi^2 \over 6} , \nonumber \\
&& \hskip -7 mm
   C^{\pm}_n(x,y) = \Li_n(-x) \pm \Li_n(-y) . 
\nonumber
\end{eqnarray}

The explicit forms for the $F^\Lead_{\lambda_1\lambda_2\lambda_3\lambda_4}$ 
appearing in eq. (\ref{FiniteRemainder}) are
\[
 F^\Lead_{++++} =  
 {1\over 2} \,,  
\]
\begin{eqnarray}
&& \hskip -7 mm
F^\Lead_{-+++}   =   
     {1\over 8}  \biggl[ - (1 - x y)  {\tilde Z} 
      + 2  \biggl( {9\over y} - 10 x \biggr)  {\tilde X}  \nonumber \\
&& + \biggl(2 + 4 {x\over y^2} - 5 {x^2\over y^2} \biggr) 
       \Xi   \biggr]
+ \Bigl\{t \leftrightarrow u \Bigr\} \,,
\nonumber
\end{eqnarray}
\begin{eqnarray}
&& \hskip -7 mm
F^\Lead_{++-+}   =   
     {1\over 8}  \biggl[ \biggl(2 + 6 {x\over y^2} - 3 {x^2\over y^2}\biggr) 
                \Xi \nonumber \\
&&        - (x-y)^2  {\tilde Z}  
       + 2  \biggl({9\over y} - 8x \biggr)  {\tilde X} \biggr]
+  \Bigl\{t \leftrightarrow u \Bigr\} \,,
\nonumber 
\hskip 2.2 cm \null 
\label{FppmpL} 
\end{eqnarray}
\begin{eqnarray}
&& \hskip -7 mm
F^\Lead_{--++}   = 
 - (x^2+y^2) \biggl[ 4 \li4(-x) + {1\over 48} Z_+^4  \nonumber \\ 
&&    + ({\tilde Y} - 3{\tilde X}) \li3(-x)  
      + \Xi \li2(-x) 
\nonumber \\ 
&&      + i {\pi\over12} Z_+^3 
            + i {\pi^3\over2} X
      - {\pi^2\over12} X^2 - {109\over 720} \pi^4 \biggr] 
\nonumber \\ 
&&    + {1\over2} x (1 - 3 y) \biggl[
          \li3(-x/y) - Z_-  \li2(-x/y) \nonumber \\
&&   - \zeta_3
       + {1\over 2}  Y  {\tilde Z} \biggr]  
       + {1\over 8}
              \biggl( 14 (x-y) - {8\over y} + {9\over y^2} \biggr) 
                     \Xi
\nonumber \\ 
&&     + {1\over 16}  ( 38 x y - 13 ) {\tilde Z}
       - {\pi^2\over12}
       - {9\over4} \biggl( {1\over y} + 2x \biggr) 
       {\tilde X} 
\nonumber \\
&&        + {1\over4} x^2 \Bigl[ Z_-^3 
                     + 3 {\tilde Y} {\tilde Z} \Bigr] 
   + {1\over 4} +  \Bigl\{ t \leftrightarrow u \Bigr\} \,,
\nonumber
\end{eqnarray}
\begin{eqnarray}
&& \hskip -7 mm
 F^\Lead_{-+-+}   =   
   - 2 {x^2+1\over y^2} \biggl[  A^-_4
       - {1\over 2} {\tilde X} A^-_3 - {1 \over 48} X^4
\nonumber \\ 
&& 
       + {\pi^2\over6} \biggl( B 
                             - {1\over2} X^2 \biggr)
     + {1\over 24} {\tilde X}^2 \Xi \biggr]    
     + {4\over 9} \pi^2 {x\over y}
\nonumber \\ 
&& 
   + 2 {3 (1-x)^2 - 2 \over y^2} \biggl[ 
            C^-_4(x,y) + \li4(-x/y)
\nonumber \\
&& 
          - {\tilde Y} A^-_3   
          + {\pi^2\over6} \biggl( \li2(-x) + {1\over2} Y^2 \biggr)
          + {1\over24} Y^4
\nonumber \\
&&
          - {1\over6} X Y^3 - {7\over360} \pi^4 \biggr]
    - {2 \over 3} 
     \biggl( 8 - x + 30 {x\over y} \biggr) 
    \biggl[ 
\nonumber \\
&&
          \li3(-y)  -\zeta_3  - {\tilde Y} 
                       \Bigl(\li2(-y) - {\pi^2 \over 6} \Bigr)  
         - {1\over2} X \Upsilon \biggr]
\nonumber \\
&&
    + {1\over 6}
     \biggl( 4 y + 27 + {42\over y} 
+ {4\over y^2} \biggr)
                 \biggl[
           i {\pi\over2} X^2
         - {\tilde X} B 
\nonumber \\ 
&& 
         + A^-_3 - \pi^2 X \biggr]  
   + {1\over 12}
   \biggl( 3 - {2\over y} - 12 {x\over y^2} \biggr) 
                {\tilde X} \Xi 
\nonumber \\ 
&& 
  + 2 \biggl( 1 + {2\over y} \biggr) 
           \Bigl( \zeta_3 - {\pi^2\over6} {\tilde Y} \Bigr)
  + {1\over 24}
   \biggl(y^2 - 24 y 
\nonumber \\ 
&& 
+ 44 - 8 {x^3\over y} \biggr) 
                        {\tilde Z}
  - {1\over 24}
   \biggl(15 - 14 {x\over y} - 48 {x\over y^2} \biggr) 
                   \Xi 
\nonumber \\
&& 
  + {1\over 24}
 \biggl( 8 {x\over y} + 60 - 24 {y\over x} + 27 { y^2\over x^2} \biggr)
                   \Upsilon
  - {1\over 3} y {\tilde X} \Upsilon 
\nonumber \\
&& 
  + {1\over 12} ( 2 x^2 - 54 x - 27 y^2 ) 
       \biggl({1\over y} {\tilde X}
            + {1\over x} {\tilde Y} \biggr) \,.
\nonumber
\end{eqnarray}

Similarly, the subleading color contributions in eqs. (\ref{FiniteRemainder})
and (\ref{QEDres}) are, 
\[
F^\Sublead_{++++} =  
- {3\over 2 }\,, 
\]
\begin{eqnarray}
&& \hskip -7 mm
F^\Sublead_{-+++} =   
    {1\over 8}  \biggl[ { x^2 + 1\over y^2 } \Xi
             + {1\over 2} (x^2 + y^2) {\tilde Z}
\nonumber \\ 
&& 
             - 4  \biggl({1\over y} - x \biggr)  {\tilde X} \biggr] 
+ \Bigl\{t \leftrightarrow u \Bigr\} \,,
\nonumber
\end{eqnarray}
\[
 F^\Sublead_{++-+} =   
 F^\Sublead_{+-++} = F^\Sublead_{+++-} = 
F^\Sublead_{-+++} \,, 
\]
\begin{eqnarray}
&& \hskip -7 mm
 F^\Sublead_{--++}   =   
 - {1 \over 4} - 2 x^2 \biggl[ C^+_4(x,y)
  - {\tilde X} C^+_3(x,y)  
\nonumber \\
&& 
  + {1\over12} X^4 - {1\over3} X^3 Y + {\pi^2 \over 12} X Y 
  - {4 \over 90} \pi^4
\nonumber \\
&&
  + i {\pi\over6} X \Bigl( X^2 - 3 X Y + \pi^2 \Bigr) \biggr] 
  + {\pi^2 \over 12}
\nonumber \\ 
&& 
-(x-y) \Bigl( \li4(-x/y) - {\pi^2\over6} \li2(-x) \Bigr) 
\nonumber\\ 
&&
 - x \biggl[ 2 \li3(-x) - \li3(-x/y) - 3 \zeta_3 
\nonumber \\ 
&&
      + Z_- ( \li2(-x/y) + X^2 ) 
   - 2 {\tilde X} \li2(-x) 
\nonumber \\
&&
   + {1\over 12} ( 5 Z_- + 18 i \pi) {\tilde Z}
   - i \pi (Y^2 + \pi^2)
\nonumber \\ 
&& 
    - {2\over 3}  X  (X^2 + \pi^2) \biggr]
    - {1\over 8} ( 2 x y + 3 ) {\tilde Z} 
\nonumber \\ 
&& 
  + { 1 - 2 x^2 \over 4 y^2 } \Xi
  + \biggl( {1\over 2 y} + x \biggr) {\tilde X}  
                    +  \Bigl\{ t \leftrightarrow u \Bigr\} \,,
\nonumber
\end{eqnarray}
\begin{eqnarray}
&& \hskip -7 mm
 F^\Sublead_{-+-+} =   
    - {1 \over 2} - 2 {x^2+1 \over y^2} \biggl[ 
       C^-_4(x/y,y)
      + {7\over 360} \pi^4
\nonumber \\
&&
    + {1\over 24} ( X^4 + 2 i \pi X^3 - 4 X Y^3 + Y^4 
              + 2 \pi^2 Y^2 ) 
\nonumber \\
&&
   + {1\over2} ({\tilde X} - 2 {\tilde Y} ) A^-_3  \biggr] 
   - 2 {x-1\over y} \biggl[ A^-_4
       - {1\over 2} {\tilde X} A^-_3
\nonumber \\
&&  
       + {\pi^2\over 6} \Bigl( B  - {1\over2} X^2 \Bigr) 
       - {1\over 48} X^4 \biggr] 
     - {2-y^2 \over 4 x^2} \Upsilon
\nonumber \\
&&
 + \biggl(2 {x\over y} - 1\biggr) 
    \biggl[ A^+_3 - {\tilde X} \li2(-x) 
             - {1\over 6} X^3
\nonumber \\
&&
 - {\pi^2\over 3} Z_+ \biggr] + {\pi^2 \over 6}
 + 2 \biggl(2 {x\over y} + 1 \biggr) 
     \biggl[ \li3(-y) 
\nonumber \\
&&
     + {\tilde Y} \li2(-x) - \zeta_3 
     + {1\over 4} X ( 2 Y^2 + \pi^2 )   
\nonumber \\
&&
     - {1\over 8} X^2 (X + 3 i\pi) \biggr]
     - {1\over 4} ( 2 x^2 - y^2 ) {\tilde Z}  
\nonumber \\
&&
     - {1\over 4} \Bigl(3 + 2 {x\over y^2} \Bigr) \Xi
     + {1\over 2} ( 2 x + y^2 ) \biggl[ {1\over y} {\tilde X} 
                                     + {1\over x} {\tilde Y} \biggr]  \,,
\nonumber
\end{eqnarray}
\[
F^\Sublead_{+--+}(s,t,u) =  
 F^\Sublead_{-+-+}(s,u,t) \,.
\nonumber
\]
The reliability of these results was ensured by performing a series of
checks described in ref.~\cite{gggamgam}.

A companion talk in these proceedings~\cite{ZviTalk} describes the
application of the amplitudes presented here to obtain~\cite{Hbkgd}
an improved prediction for the QCD background to Higgs production at the
LHC, when the Higgs decays into two photons.


\end{document}